\documentclass[superscriptaddress,floatfix,aps,pre,twocolumn,notitlepage,shortbibliography,nofootinbib]{revtex4-1}
\usepackage[utf8]{inputenc}
\usepackage{amsmath}
\usepackage{amssymb}
\usepackage{refcount}
\usepackage{siunitx}
\usepackage{graphicx}
\usepackage{hyperref}
 \usepackage[capitalise]{cleveref}
\usepackage{microtype}
\usepackage{bm}
\usepackage[english]{babel}
\usepackage{color}
\usepackage{graphicx}
\usepackage{xcolor}
\usepackage{braket}

\newcommand{\comm}[2]{\left[ #1 , #2 \right]}
\newcommand{\acomm}[2]{\left\{ #1, #2 \right\}}

\newcommand{\br}{\mathbf{r}}

\newcommand{\bp}{\mathbf{p}}

\newcommand{\bE}{\mathbf{E}}
\newcommand{\bB}{\mathbf{B}}
\newcommand{\bA}{\mathbf{A}}

\newcommand{\balpha}{\bm\alpha}

\begin{document}
\author{Haidar Al-Naseri}
\email{haidar.al-naseri@umu.se}
\affiliation{Department of Physics, Ume{\aa} University, SE--901 87 Ume{\aa}, Sweden}

\author{Jens Zamanian}
\author{Gert Brodin}
\email{gert.brodin@umu.se}
\affiliation{Department of Physics, Ume{\aa} University, SE--901 87 Ume{\aa}, Sweden}
\title{Spacetime-dependent electric field effects in vacuum and plasma using the Wigner-formalism }

\begin{abstract}
We derive a system of coupled partial differential equations for the equal-time Wigner function in an arbitrary strong electromagnetic field using the Dirac-Heisenberg-Wigner formalism. In the electrostatic limit, we present a 3+1-system of four coupled partial differential equations, which are completed by Ampères law. This electrostatic system is further studied for two different cases. In the first case, we consider linearized wave propagation in plasma accounting for the nonzero vacuum expectation values. We then derive the dispersion relation and compare it with well-known limiting cases.  
In the second case, we consider Schwinger pair production using the local density approximation to allow for analytical treatment. The dependence of the pair production rate on the perpendicular momentum is investigated and it turns out that the spread of the produced pairs along with perpendicular momentum depends on the strength of the applied electric field.
   
\end{abstract}
 
\maketitle

\section{Introduction}
Quantum relativistic treatment of plasmas are of interest in several
different contexts \cite{QRP-1,QRP-2,QRP-3}. Dense astrophysical objects can have a
Fermi energy approaching or exceeding the electron rest mass energy, the
strong magnetic fields of magnetars give raise to relativistic Landau
quantization, and the high plasma density in the early universe imply yet
new phenomena. In the laboratory, the continuous evolution of laser
intensity brings a variety of quantum relativistic phenomena accessible to
experimentalists. Upcoming laser facilities of interest in this context
includes e.g. the extreme light infrastructure (ELI) \cite{Eli,Dunne} and
the European x-ray free electron laser (XFEL) \cite{XFEL,Ringwald}, that
will facilitate experimental observations of various fundamental processes.
Already with existing technology, laser-induced spin polarization seems
possible \cite{SP-1,SP-2,SP-3}. Moreover, radiation reaction might take place at least
partially in the quantum relativistic regime \cite{QRR}. A particular phenomena of
much interest is electron-positron pair production \cite{Gies,Gies 2,Florian,Kohlfurst,Kohlfurst-2020,Sheng,Vasak,Bloch}, that has received much
attention  since this interesting process might eventually be viable in
the laboratory. 

Simplified quantum relativistic models of plasmas have been presented by 
e.g. \cite{Asenjo,Manfredi}, focusing on the weakly
relativistic regime. Extensions to the strongly relativistic regime has been
made by e.g. Ref. \cite{Ekman,Ekman2}, although certain simplifying
assumptions have been made concerning e.g. the scale lengths of interest.
However, quantum kinetic relativistic model based of the full Dirac equation
are derived in \cite{Bloch,Vasak-87,Kluger,Smolyansky,Birula}. While these equations
are applicable to plasma dynamics in general, much of the analysis of these
models have been devoted to the phenomena of pair-production in vacuum by
high-intensity fields due to the Schwinger mechanism \cite{Sauter,Schwinger}.

In the present paper, we will adopt the Dirac-Heisenberg-Wigner (DHW)
formalism of Ref. \cite{Birula} and apply it to electrostatic phenomena in
plasmas and vacuum. Specifically, we will reduce the general DHW-system to 4
coupled equations, in the limit of 1D spatial variations. The simplified
system is used to derive a dispersion relation for Langmuir waves,
demonstrating that wave-particle interaction with the quantum vacuum is
possible, leading to electron-positron pair-creation. Moreover, the reduced
electrostatic equations are used to study the influence of perpendicular
momentum (perpendicular referring to the direction of the electric field) on
the process of pair production in vacuum. While the common omission of
perpendicular momentum can be justified to some degree, we point out some
significant corrections introduced by incorporating the full momentum
dependence. Finally, we present our main conclusions and provide and outlook
for future work.

\section{The DHW-formalism}
\label{DHW-formalism}
In this section a brief review of the DHW-formalism of Ref. \cite{Birula} is given. The theory is then applied to the case of one-dimensional electrostatic fields. In this limit, the full set of 16 scalar DHW-functions is reduced to four scalar equations, which form a self-consistent system together with Ampere's law.  

\subsection{DHW equation of motion}
In this subsection, we derive a set of expansion coefficients, which we term the DHW-functions, of the equal-time Wigner operator $\hat{W}(\br,\bp, t)$. We use the temporal gauge  where the scalar potential $\phi$ is set to zero, thus the electromagnetic field is given  by $\bE=-\partial_t \bA$ and $\bB= \nabla \times \bA$. 
The gauge-fixing slightly simplifies the derivation of the evolution equations for the DHW-functions. However, since a gauge-independent Wigner transformation is utilized, the end result will be gauge-invariant. 

Our starting point is the Dirac equation in the temporal gauge
\begin{equation}
\label{Dirac_eq_1}
    \Big[i\partial_t +\balpha \cdot (i\bm\nabla + e\bA)+\beta m \Big] \hat{\Psi}(\br, t)=0 .
\end{equation}
We use the gauge independent Wigner transformation 
\begin{multline} \label{wigtrans}
\hat{W}(\br,\bp) = 
\\
\int d^3z
\exp \bigg(-i\textbf{p}\cdot\textbf{z} -ie\int^{1/2}_{-1/2} d\lambda \mathbf{z}\cdot \mathbf{A}(   \br+\lambda\mathbf{z},t )\bigg) 
\\ 
\times \hat{C}(\br,\bp,t),
\end{multline}
where
\begin{equation}
\label{Dirac-operatos}
    \hat{C}(\br,\bp,t) = - \frac{1}{2}
\comm{ \hat{\Psi}(\br+\mathbf{z} /2,t )} { \hat{\bar{\Psi}} (\br -\mathbf{z}/2,t )} . 
\end{equation}
In \cref{wigtrans} we use the Wilson line factor to ensure the gauge invariance.
The Wigner function $W(\br,\bp,t)$ is defined as the expectation value of the Wigner operator
\begin{equation}
\label{Wigner_function}
    W(\br,\bp,t)=\braket{\Omega |\hat{W}(\br,\bp,t)|\Omega}, 
\end{equation}
where $\ket{\Omega}\bra{\Omega}$ is the state of the system. 
In order to derive an equation of motion for the Wigner function, we take the time derivative of \cref{Wigner_function}. 
We use the Hartree approximation where the electromagnetic field is treated as a non-quantized field. This approximation is well justified for high electromagnetic field strengths and amounts to neglecting the quantum fluctuations. 
Applying the Hartree approximation we replace 
\begin{align}
\label{hartree}
    \braket{\Omega|\bE(\br,t) \hat{C}(\br,\bp,t) |\Omega} &\rightarrow\braket{\Omega|\bE(\br,t) |\Omega} \braket{\Omega| \hat{C}(\br,\bp,t) |\Omega} \notag  \\
     \braket{\Omega|\bB(\br,t) \hat{C}(\br,\bp,t) |\Omega} &\rightarrow\bra{\Omega|\bB(\br,t) |\Omega} \braket{\Omega| \hat{C}(\br,\bp,t) |\Omega}
\end{align}
This approximation corresponds to ignoring higher-loop radiative corrections and is appropriate for fields that varies slowly with time \cite{Temporal}. Finally, the equation of motion of the Wigner function is given by \cite{Birula}
\begin{equation}
\label{Kinetic_equation}
    i\hbar D_t W(\br,\bp) =m \comm{\beta}{W}
    + \comm{\Tilde{\textbf{p}}\cdot \balpha }{W}
    -\frac{i\hbar}{2} \acomm{\textbf{D}}{W},
\end{equation}
where we have the non-local operators
\begin{align}
D_t&= \frac{\partial}{\partial t} + e\Tilde{\bE}\cdot \bm\nabla_p\\
    \Tilde{\textbf{p}}&=\textbf{p}- ie\int^{1}_{-1}d\tau \tau \textbf{B}(\br+i\hbar \tau \bm \nabla_p)\times \bm\nabla_p\\
    \textbf{D}&= \nabla_r+ e\int^{1}_{-1}d\tau \tau \textbf{B}(\br+i\hbar\tau \bm \nabla_p)\times \bm \nabla_p\\
    \Tilde{\textbf{E}}&= \int^{1}_{-1}d\tau  \textbf{E}(\br+i\hbar \tau \bm\nabla_p) 
\end{align}
which reduce to their local approximations (i.e. $D_t\rightarrow \partial/\partial t+e{\bE}\cdot\nabla_p$ and $\Tilde{\bE}\rightarrow{\bE}$, etc.) for scale lengths much longer than the characteristic de Broglie length. 

\subsection{The DHW-expansion}
Even though the equation of motion of the Wigner function \cref{Kinetic_equation} has only a couple of terms, it is not simple to interpret it since the particle and anti-particle states are mixed. However, expanding the Wigner function $W(\br,\bp,t)$ in terms of an irreducible set of $4 \times 4$ matrices $\{\textbf{1},\gamma_5,\gamma^{\mu},\gamma^{\mu}\gamma_5,\sigma^{\mu,\nu} \}$ where $\textbf{1}$ is a $4\times 4$-identity matrix, we get
\begin{equation}
\label{Expansion}
    W(\br,\bp,t)= \frac{1}{4}\Big[ s+ i\gamma_5 \varrho + \gamma^{\mu} v_{\mu}+ \gamma^{\mu}\gamma^5 a_{\mu} + \sigma^{\mu \nu }t_{\mu \nu}
    \Big],
\end{equation}
 where the expansion coefficients $\{s,\varrho,v_{\mu},a_{\mu},t_{\mu\nu}\}$ are called the DHW-functions. 
This expansion leads to a number of coupled differential equations. The tensor part $\sigma^{\mu \nu}$ in \cref{Expansion} can be decomposed into 

\begin{equation}
    \mathbf{t}_1=\begin{pmatrix}
    t^{10}\\
    t^{20}\\
    t^{30}
    \end{pmatrix}
    , 
    \mathbf{t}_2=\begin{pmatrix}
    t^{23}\\
    t^{31}\\
    t^{12}
    \end{pmatrix}
\end{equation}
Using the expansion in \cref{Expansion} in \cref{Kinetic_equation}, and comparing the coefficients of the basis matrices, we get the following system of partial differential equations
\begin{align}
\label{DHW_System_diff}
    D_t s-2\Tilde{\textbf{p}}\cdot \textbf{t}_1&=0 \notag\\
    D_t\varrho+2 \Tilde{\textbf{p}}\cdot \textbf{t}_2 &=2ma_0\notag\\
    D_tv_0 +\textbf{D} \cdot \textbf{v}&=0\notag \\
    D_ta_0 + \textbf{D}\cdot \textbf{a}&=-2m \varrho  \\
    D_t \textbf{v} + \textbf{D}v_0 - 2\Tilde{\textbf{p}}\times \textbf{a}&=-2m\textbf{t}_1\notag \\
    D_t\textbf{a} + \textbf{D}a_0 - 2\Tilde{\textbf{p}}\times \textbf{v}&=0\notag \\
    D_t\textbf{t}_1+ \textbf{D}\times \textbf{t}_2 + 2\Tilde{\textbf{p}} s&=2m\textbf{v} \notag \\
    D_t \textbf{t}_2- \textbf{D}\times \textbf{t}_1 
    -2\Tilde{\textbf{p}}\varrho &=0.\notag 
\end{align}
Thus we have 16 scalar components of coupled partial differential equations. This system can be expressed in matrix-form as
\begin{equation}
\label{DHW_System}
    D_t
    \begin{pmatrix}
    G_1\\
    G_2\\
    G_3\\
    G_4
    \end{pmatrix}
    =
    \begin{pmatrix}
    0&0&0& M_1\\
    0&0&-M_2&0\\
    0&-M_2&0&-2m\\
    -M_1&0&2m&0
    \end{pmatrix}
       \begin{pmatrix}
    G_1\\
    G_2\\
    G_3\\
    G_4
    \end{pmatrix},
\end{equation}
where we have divided the DHW-functions into four groups
\begin{align}
    G_1=
    \begin{pmatrix}
    s\\
    \mathbf{t}_2
    \end{pmatrix}
    ,  G_2=
    \begin{pmatrix}
    v_0\\
    \mathbf{a}
    \end{pmatrix} \notag
    \\
     G_3=
    \begin{pmatrix}
    a_0\\
    \mathbf{v}
    \end{pmatrix} 
    , G_4=
    \begin{pmatrix}
    \varrho\\
    \mathbf{t}_1
    \end{pmatrix} 
\end{align}
and we have defined
\begin{align}
  M_1=
  \begin{pmatrix}
  \textbf{0} & 2\Tilde{\mathbf{p} }\\
  2\Tilde{\mathbf{p}}& \mathbf{D}^{x}
  \end{pmatrix}
  ,  M_2=
  \begin{pmatrix}
  \textbf{0} & \mathbf{D}\\
  \mathbf{D} & -2\Tilde{\mathbf{p}}^x
  \end{pmatrix}
\end{align}
where $\mathbf{D}^x$ is the anti-symmetric representation of $\mathbf{D}$.

One can show that some of the DHW-functions have a clear physical interpretation. Firstly, the electromagnetic current $J^{\mu}$ can be expressed
\begin{equation}
J^{\mu}= \frac{e}{(2\pi)^3} \int d^3p\, v^{\mu} (\br,\bp,t)
\end{equation}
where the total charge Q is
\begin{equation}
\label{Conservation_charge}
    Q=\frac{e}{(2\pi)^2}\int d^3pd^3x  v_0(\br,\bp,t)
\end{equation}
Moreover, the total energy $W$ is given by
\begin{multline}
\label{Conservation_Energy}
W= \frac{1}{(2\pi)^3}\int d^3pd^3x \big[\mathbf{p}\cdot \mathbf{v}(\br,\bp,t) + ms(\br,\bp.t)   \big]\\
+ \frac{1}{2} \int d^3x \big[E^2+B^2   \big].
\end{multline}
The linear momentum is 
\begin{equation}
\label{Momentum}
    \textbf{p}= \frac{1}{(2\pi)^2}\int d^3pd^3x\, \textbf{p} v_0(\br,\bp,t)  + \int d^3x \textbf{E}\times \textbf{B}
\end{equation}
and the total angular momentum $\textbf{M}$ is 
\begin{multline}
\label{Angular_Mom}
    \textbf{M}= \frac{1}{(2\pi)^2}\int d^3pd^3r \Big[\textbf{r}\times \textbf{p} v_0(\br,\bp,t) + \frac{1}{2}\textbf{a}(\br,\bp,t)  \Big]\\ + \int d^3r \, \textbf{r}\times \textbf{E}\times \textbf{B}
\end{multline}
The interpretation that can be done from the expressions above that $s(\br,\bp,t)$ is the mass density, $v_0(\br,\bp,t)$ is the charge density and $\mathbf{v}(\br,\bp,t)$ is the current density. Moreover, the function $\mathbf{a}$ can be associated with the spin density. 

The classical, but still relativistic, Vlasov equation can be obtained by in the limit $\hbar \rightarrow 0$. 
Note, however, that the variable $v_0$, which is proportional to the charge density, must be kept non-zero. 
Thus the procedure to reach the classical limit, which is outlined in Ref.~\cite{Birula}, must be somewhat modified. 


\subsection{Space and time-dependent electrostatic fields}
In this subsection, we simplify the DHW-system \cref{DHW_System} by considering one-dimensional electrostatic fields, $\bE(t,\br)=E(t,z)\textbf{e}_z$. This simplifies the operators $M_1$ and $M_2$ to
\begin{align}
  M_1=
  \begin{pmatrix}
  \textbf{0} & 2\mathbf{p} \\
  2\mathbf{p}& \bm{\nabla}^{x}
  \end{pmatrix}
  ,  M_2=
  \begin{pmatrix}
  \textbf{0} & \bm{\nabla}\\
  \bm{\nabla} & -2\mathbf{p}^x
  \end{pmatrix}.
\end{align}
By considering an electrostatic geometry, we got rid of complicated operators that depend on the magnetic field. However, we still have 16 coupled scalar-functions, which we can expand as
\begin{equation}
    G(z,\bp,t)=\{G_1,G_2,G_3,G_4\}=\sum_{i=1}^{16} \chi^i(z,\bp,t)\textbf{e}_i(z,\bp,t),
\end{equation}
where $\chi^{i} (z,\bp,t)$ are expansion coefficients and $\textbf{e}_i(z,\bp,t)$ are orthonormal basis vectors.
Since $G(z,\bp,t)$ is a $16$-vector, we need  a set of 16 unit vectors.  Sheng et al. \cite{Sheng} considered basis vectors that only depended on $\textbf{p}_{\bot}$ for the case of a homogeneous electric field. The point of having such basis is that the they will not be acted on by the operator $D_t$ and hence one can close the system in a less complicated way. In order to close the system for the homogeneous field case, Sheng et al. used three basis vectors. However, since we consider a space time-dependent electric field, it turns out we need to define one more unit vector. As we will see, we can express $G(z,\bp,t)$ as
\begin{equation}
\label{Ansatz_perp}
    G(z,\bp,t)=\sum_{i=1}^{4} \chi^i(z,\bp,t)\textbf{e}_i(p_{\bot})
\end{equation}
with the four basis vectors
\begin{align}
    \textbf{e}_1&=
    \begin{pmatrix}
    0\\
    0\\
   \begin{pmatrix}
    0\\
      \textbf{e}_z
      \end{pmatrix}
      \\
    0
    \end{pmatrix}
    ,\, 
    \textbf{e}_2=\frac{1}{\epsilon_{\bot}}
       \begin{pmatrix}
       \begin{pmatrix}
    m\\
    \mathbf{0}
    \end{pmatrix}
    \\
    0\\
    \begin{pmatrix}
    0\\
    \textbf{p}_{\bot}
\end{pmatrix}
    \\
    0
    \end{pmatrix}\notag
    \\
    \textbf{e}_3&= \frac{1}{\epsilon_{\bot}}
       \begin{pmatrix}
    0\\
    \begin{pmatrix}
    0\\
    \textbf{e}_z\times \textbf{p}_{\bot}
    \end{pmatrix}
    \\
    0\\
    \begin{pmatrix}
    0\\
    -m\textbf{e}_z
    \end{pmatrix}
    \end{pmatrix}
    , \, \textbf{e}_4=
       \begin{pmatrix}
    0\\
    \begin{pmatrix}
    1\\
    \mathbf{0} 
    \end{pmatrix}
    \\
0\\
    0
    \end{pmatrix}
\end{align}
where $\epsilon_{\bot}=\sqrt{m^2+p_{\bot}^2}$.
Note that $\textbf{e}_4$ is the extra basis vector that we need to define in order to close the system in our case.  Using \cref{Ansatz_perp} in \cref{DHW_System}, we finally get
\begin{align}
\label{PDE_System}
    D_t\chi_1(z,\bp,t)&= 2\epsilon_{\bot}(p_{\bot}) \chi_3(z,\bp,t)- \frac{\partial \chi_4}{\partial z} (z,\bp,t)\notag\\
    D_t\chi_2(z,\bp,t) &= -2p_z\chi_3(z,\bp,t)\\
    D_t\chi_3(z,\bp,t)&= -2\epsilon_{\bot}(p_{\bot}) \chi_1(z,\bp,t) +2p_z\chi_2(z,\bp,t)\notag\\
    D_t\chi_4(z,\bp,t)&= -\frac{\partial \chi_1}{\partial z}(z,\bp,t)\notag 
\end{align}
This system of four coupled equations is closed by Ampére's law
\begin{equation}
\label{Ampers_law}
\frac{\partial E}{\partial t}=e\int \chi_1 d^3p
\end{equation}
where we have used the relation between the original DHW-functions and the the expansion functions $\chi_i(z,\bp,t)$. The complete list of relations between the variables are as follows:
\begin{align}
\label{DHW_Non}
    s(z,\bp,t)&= \frac{m}{\epsilon_{\bot}}\chi_2(z,\bp,t) \notag\\
    v_0(z,\bp,t)&= \chi_4(z,\bp,t)\notag \\
    \textbf {v}_{\bot}(z,\bp,t)&=\frac{ \textbf{p}_{\bot}}{\epsilon_{\bot}}\chi_2(z,\bp,t) \notag \\
    v_z(z,\bp,t)&=\chi_1(z,\bp,t) \\
    a_x(z,\bp,t)&=-\frac{p_y}{\epsilon_{\bot}}\chi_3(z,\bp,t)\notag\
    a_y(z,\bp,t)&=\frac{p_x}{\epsilon_{\bot}}\chi_3(z,\bp,t)\notag\\
    t_{1z}(z,\bp,t)&=- \frac{m}{\epsilon_{\bot}} \chi_3(z,\bp,t) \notag 
\end{align}
As seen above, for the electrostatic case of consideration we have 8 non-zero DHW-functions. The PDE-system in \cref{PDE_System} can be verified by using the relations between these 8 DHW-functions in the general system of \cref{DHW_System_diff}.

\section{Linear waves}
\label{Linear-waves}
In this section, we will demonstrate the usefulness of \cref{PDE_System}
and (\ref{Ampers_law}) by considering linearized wave propagation in
plasmas, accounting also for the contribution from the nonzero vacuum
background expectation values. For our case with no background
electromagnetic fields, we get the unperturbed vacuum contributions as the
Wigner transform of the expectation value of the free Dirac field operators.
Forgetting about the contribution from real electrons and positrons
to start with, we note that the only nonzero DHW-functions in the vacuum
background are 
\begin{align}
s_\text{vac}(\bp)& =-\frac{2m}{\epsilon }  \notag  \label{Vacuum_sol} \\
\mathbf{v}_\text{vac}(\bp)& =-\frac{2\mathbf{p}}{\epsilon },
\end{align}%
where $\epsilon=\sqrt{m^2+\bp^2}$.
The expressions above are obtained by calculating the Wigner operator for the free particle Dirac equation and taking the vacuum expectation value. 
The nonzero vacuum contributions to the functions $\chi _{i}
$ become 
\begin{align}
\chi _{1}(\bp)& =-\frac{2p_{z}}{\epsilon }  \notag \\
\chi _{2}(\bp)& =-\frac{2\epsilon _{\bot }}{\epsilon }.
\end{align}%
A background distribution
function $f_{e}(\bp)$ of electrons ($f_{p}(\bp)$ for positrons), normalized
such that the unperturbed number density $n_{0}$ is
\begin{equation}
    n_{0}=\frac{2}{(2\pi \hbar
)^{3}}\int f_{e,p}(\bp)d^{3}p,
\end{equation}
can be added to the vacuum background as
follows: 
\begin{align}
v_{0}& =2(F+1) \\
s(\bp)& =\frac{2m}{\epsilon }F(\bp) \\
\mathbf{v}(\bp)& =\frac{2\mathbf{p}}{\epsilon }F(\bp),
\end{align}%
where $F(\bp)=[f_{p}(\bp)+f_{e}(\bp)-1]$. Here $f_{p/e}(\bp)$ can be picked as any common background distribution function from classical kinetic theory, i.e. a Maxwell-Boltzmann, Synge-Juttner, or Fermi-Dirac distribution, depending on whether the characteristic kinetic energy is relativistic and whether the particles are degenerate. 

Note that for a
completely degenerate ($T=0$) Fermi-Dirac background of electrons (and no positrons $f_{p}=0$), the electron and vacuum contributions cancel  inside the Fermi sphere. Consequently, for
momenta $p\leq p_{F}$, where  $p_{F}=\hbar (3\pi^2n_{0})^{1/3}$ is the Fermi
momentum we have $F(\bp)=0$. In terms of the functions $\chi _{i}$, we have 
\begin{align}
\chi _{1}^{0}(\bp)& =\frac{2p_{z}}{\epsilon }\Big[f_{p}(\bp)+f_{e}(\bp)-1\Big] 
\notag \\
\chi _{2}^{0}(\bp)& =\frac{2\epsilon _{\bot }}{\epsilon }\Big[%
f_{p}(\bp)+f_{e}(\bp)-1\Big] \\
\chi _{4}^{0}(\bp)& =2\Big[f_{p}(\bp)-f_{e}(\bp)\Big]  \notag
\end{align}%
using upper index $0$ for the unperturbed background values. 
Next, we divide the variables into unperturbed and perturbed variables
according to 
\begin{equation}
\chi _{i}(z,\bp,t)=\chi _{i}^{0}(\bp)+\chi _{i}^{1}(\bp)e^{i(kz-\omega t)}
\end{equation}%
(with $\chi _{3}^{0}(\bp)=0$ and only a perturbed electric field $E$) and linearize
\cref{PDE_System} and (\ref{Ampers_law}). Making use of the
relation 
\begin{equation}
\Tilde{\mathbf{E}}\cdot \nabla _{p}\chi _{i}^{0}=\tilde{E}\frac{\partial
\chi _{i}^{0}}{\partial p_{z}}=E\frac{\chi _{i}^{0}(p_{z}+\hbar k/2)-\chi
_{i}^{0}(p_{z}-\hbar k/2)}{\hbar k}  \label{Derivative generalization}
\end{equation}%
the problem is reduced to linear algebra.  Solving for $\chi _{i}^{1}(\bp)$ we
obtain
\begin{widetext}

\begin{align}
    \chi_1(\bp)&=   \sum_{\pm}
    \frac{\pm i2e\omega E/ (\hbar k) }{(\omega^2-k^2)(\hbar^2\omega^2-4p_z^2)-4\epsilon_{\bot}^2\omega^2}
    \Bigg[ 4p_z\epsilon_{\bot}^2   \frac{F(p_{\pm})}{\epsilon_{\pm}}  
    -(\hbar^2\omega^2 -4p_z^2)
    \bigg( \frac{p_{\pm}}{\epsilon_{\pm}} F(p_{\pm})
  + \frac{k}{\omega}\Big(f_p(p_{\pm})-f_e(p_{\pm})\Big)
    \bigg)
    \Bigg]
    \label{chi1} 
    \\
    \chi_2(\bp)&=  \sum_{\pm}
    \frac{\mp  i\omega eE \epsilon_{\bot } / (\hbar k) }{(\omega^2-k^2)(\hbar^2\omega^2-4p_z^2)-4\epsilon_{\bot}^2\omega^2}
    \Bigg[  \Big( \hbar^2 \omega^2 -\hbar ^2k^2 - 4\epsilon^2 \mp \frac{\hbar k }{2}p_z\Big)   \frac{F(p_{\pm})}{\epsilon_{\pm}}  
  - 4p_z\frac{k}{\omega}\Big(f_p(p_{\pm})-f_e(p_{\pm})\Big)
    \Bigg]\\
    \chi_3(\bp)&= \sum_{\pm}
    \frac{\mp  4\omega eE \epsilon_{\bot }  }{(\omega^2-k^2)(\hbar^2\omega^2-4p_z^2)-4\epsilon_{\bot}^2\omega^2}
    \Bigg[  \Big( p_z\frac{k}{\omega}\pm \frac{\hbar \omega }{2}\Big)\frac{F(p_{\pm})}{\epsilon_{\pm}}   + f_p(p_{\pm})-f_e(p_{\pm})
    \Bigg]\\
    \chi_4(\bp)&= \sum_{\pm}
    \frac{\pm  2i\omega eE/(\hbar k)  }{(\omega^2-k^2)(\hbar^2\omega^2-4p_z^2)-4\epsilon_{\bot}^2\omega^2}
    \Bigg[  \big(4\epsilon^2-\hbar^2\omega^2  \big)
    \left[
    \frac{kp_z}{\omega}\frac{F(p_{\pm})}{\epsilon_{\pm}}  + f_p(p_{\pm})-f_e(p_{\pm})\right] 
    \pm \frac{\hbar k^2}{2\omega}
    \big(4p_z^2-\hbar ^2\omega^2   \big) \frac{F(p_{\pm})}{\epsilon_{\pm}} 
    \Bigg]
    \label{chi4}
\end{align}

where
\begin{align}
p_{\pm}&= p_z\pm \frac{\hbar k}{2}\\
\epsilon_{\pm}&= \sqrt{m^2+p_{\bot}^2 + \Big(p_z \pm \frac{\hbar k }{2}\Big)^2}
\end{align}
Note that $F(p_{\pm})$ and $f_{e,p}(p_{\pm})$ depend on the full momentum, but we suppressed the perpendicular momentum to simplify the notation.
Combining the above results for $\chi_i(\bp)$ with Ampere's law \cref{Ampers_law} we obtain the dispersion relation $D(k,\omega)=0$ with
\begin{multline}
    D(k,\omega)= 1+ \sum_{\pm} \int d^3p
    \frac{\pm 2e^2 / (\hbar k) }{(\omega^2-k^2)(\hbar^2\omega^2-4p_{\pm}^2)-4\epsilon_{\bot}^2\omega^2}
    \Bigg[ 4\frac{\epsilon_{\bot}^2}{\epsilon}   p_{\pm} F(\bp)  
    -(\hbar^2\omega^2 -4p_{\pm}^2)
    \bigg( \frac{p_z}{\epsilon } F(\bp)
  + \frac{k}{\omega}\Big(f_p(\bp)-f_e(\bp)\Big)
    \bigg)
    \Bigg] \label{Full-DR}
\end{multline}
The classical, but relativistic, limit of the dispersion relation is obtained by letting $\hbar \rightarrow 0$. Taking this limit, the dispersion function (\ref{Full-DR}) reduces to
\begin{equation}
    D(k,\omega )= 1+ \frac{e^2}{\omega } \int d^3p \frac{p_z}{\epsilon}
    \bigg(\frac{1}{\omega-kp_z/\epsilon}+ \frac{1}{\omega+kp_z/\epsilon}
    \bigg)
    \bigg[ \Big(1 + \frac{kp_z}{\epsilon \omega}
    \Big)\frac{\partial f_p(\bp)}{\partial p_z}
    + \Big(1 - \frac{kp_z}{\epsilon \omega}
    \Big)\frac{\partial f_e(\bp)}{\partial p_z}
    \bigg], 
\end{equation}
\end{widetext}
which can be shown to agree with the standard result after some straightforward algebra. 

The main purpose of this section has been to demonstrate the usefulness of
\cref{PDE_System,Ampers_law} to problems in plasma physics, including effects due to the vacuum background. 
However, the quantum relativistic generalization of Langmuir waves is
interesting in its own right, and the full dispersion function (\ref{Full-DR}) will be
thoroughly investigated in a forthcoming paper. Here the vacuum polarization
contribution to (\ref{Full-DR}) will be of much interest, and also the issue of pair-production, as induced by wave-particle interaction with the quantum vacuum.  
As it turns out, a complete treatment of the quantum vacuum will require a renormalization, in order to remove the ultra-violet divergences \cite{Birula}, i.e. the high momentum divergences in the integrals \crefrange{chi1}{chi4}. These divergences are of logarithmic type.

\section{Schwinger Pair-Production }
\label{Schwinger-Pair-prod.}
Next, we will abandon the simplifying assumption of linearized theory, and allow for an electric field of arbitrary strength, in order to study Schwinger pair-production. To simplify matters, and allow for an analytical treatment we will make two simplifying assumptions. Firstly, we will consider a pure vacuum initially, and secondly, we will not solve for the electrostatic field self-consistently (using Ampere's law), but instead consider the response to a prescribed pulse, localized in space and time.       
\subsection{Pair-production rate}
To derive an expression for the number of produced pairs, we can make us of the conservation of energy in \cref{Conservation_Energy}. By requiring that the total energy of particles is
\begin{equation}
    W= \int d^3p d^3x \, \epsilon(\bp)\, n(z,\bp,t)
\end{equation}
where $n(z,\bp,t)$ is the number particle density, we get
\begin{equation}
    n(z,\bp,t)= \frac{m}{\epsilon} s(z,\bp,t)+ \frac{\textbf{p}}{\epsilon}\cdot \textbf{v}(z,\bp,t).
\end{equation}
Hence, the number of produced particles due to the prescribed electric field is 
\begin{equation}
\label{Pair_Prod}
    n(z,\bp,t)=\frac{m}{\epsilon} \Big[ s(z,\bp,t)- s_{i}(\bp)\Big]+ \frac{\textbf{p}}{\epsilon}\cdot    \Big[\textbf{v}(z,\bp,t) - \textbf{v}_{i}(\bp)\Big]  
\end{equation}
where $s_{i}$ and $\textbf{v}_{i}$ are the mass and current density initially. 
Assuming that we have vacuum before the electric pulse appears, we can use \cref{Vacuum_sol} for these initial values and \cref{Pair_Prod} reduces to
\begin{equation}
\label{Pair_Prod2}
    n(z,\bp,t)=2+\frac{1}{\epsilon} \Big[m s(z,\bp,t)+ \bp \cdot \textbf{v} (z,\bp,t)\Big]
\end{equation}
Next we want now to utilize \cref{PDE_System} and \cref{DHW_Non}, to simplify the expression for the number of pairs $n(z,\bp,t)$. After some algebra
 \cref{PDE_System} and \cref{DHW_Non} gives us the following relation 
\begin{equation}
    p_zs(z,\bp,t)= \frac{m}{4\epsilon_{\bot}^2} D_t\Big[D_t v_z + \frac{\partial v_0}{\partial z}\Big]+ m v_z.
\end{equation}
This can be used in 
\cref{Pair_Prod2} to express the number of pairs $n(z,\bp,t)$ in terms of the current density $v_z(z,\bp,t)$ and the charge density $v_0(z,\bp,t)$. Performing this final step, we get
\begin{multline}
\label{Number_of_pairs}
    \Tilde{n}(z,\bp,t)= 2p_z + \bigg[\epsilon + \frac{1}{4\epsilon} \Big(D_t^2 - \frac{\partial^2}{\partial z^2} \Big) \bigg]v_z(z,\bp,t)\\- \frac{e}{4\epsilon} \frac{\partial E}{\partial z}\frac{\partial v_0}{\partial p_z}
\end{multline}
where we have introduced $\Tilde{n}(z,\bp,t)=p_zn(z,\bp,t)$. 

In the next subsection, we will study the number of pairs expressed in \cref{Number_of_pairs} using the local density approximation.
\subsection{Local density approximation}
For an electric field that is given in the form 
\begin{equation}
    E(z,t)=E_0g(t)f(z)
\end{equation}
and assuming that the spatial variation of the electric field is much longer than the Compton wavelength $\lambda \gg \lambda_c$, it is possible to describe the Schwinger effect at any point $z_\text{fix}$ independently. Our goal is to use the analytical solution of the one-particle distribution function $F(\bp,t)$ for a homogeneous electric field \cite{Kluger,Smolyansky,Gies}. Thus, we approximate the current density $v_z(\bp,t,z)$ as
\begin{equation}
    v_z(\bp,t,z)\approx v_{z}^{h}\Big(\bp,t,E_0f(z)\Big)
\end{equation}
where $v_{z}^{h}(\bp,t,E_0f(z))$ is the current density from the analytical solution of the homogeneous case where $E(t)$ has been replaced by $E(t) f(z_\text{fix})$. Thus, the number of produced pairs in local density approximation is
\begin{multline}
\label{n_loc}
    \Tilde{n}_\text{loc}(\bp,t)= \int dz \Bigg( 2p_z + \bigg[\epsilon + \frac{1}{4\epsilon} \Big(D_t^2 - \frac{\partial^2}{\partial z^2} \Big) \bigg]v_z^h\Big(\bp,t,E_0f(z)\Big)\\- \frac{e}{4\epsilon} \frac{\partial E}{\partial z}\frac{\partial v_0}{\partial p_z} \Bigg)
\end{multline}

For a spatially and temporally well-localized pulse, the electric field is ideally given by
\begin{equation}
\label{Electric_field}
    E(z,t)=E_0 \exp\Big(-\frac{z^2}{2\lambda^2}\Big) \rm{sech}^2\Big(\frac{t}{\tau}\Big)
\end{equation}
where $\tau$ is the time duration of the pulse. We are interested in studying the number of produced pairs at a time when the electric field has vanished. This is because the interpretation of $\Tilde{n}_\text{loc}(p,t)$ as the momentum distribution of real particles is not sharply well defined until we take the asymptotic limit $t\rightarrow \infty$.  
Moreover, the analytical expression of $v^h_z\Big(\bp,t,E_0f(z)\Big)$ becomes much simplified when we take the limit $t\rightarrow \infty$.
By taking the asymptotic limit, we note that
the third term in $\Tilde{n}_\text{loc}(p,t)$ vanishes. However, we need to 
calculate the operators that are acting on $v_z^h(p,t,E_0f(z))$ in the second term of \cref{n_loc} before we take the limit of $t\rightarrow \infty $. We then get
\begin{equation}
\label{Local_approx}
    \Tilde{n}_\text{loc}(\bp,t\rightarrow \infty)= 2p_z\int dz F\Big(\bp,E_0f(z),t\rightarrow \infty\Big),
\end{equation}
where
\begin{widetext}
\begin{equation}
\label{Quantum_Vlasov}
    F\Big(\bp,E_0f(z),t\rightarrow \infty\Big)=\frac{
2 \sinh \Big( \frac{\pi \tau  }{2} [2\tau eE_0f(z) + \Tilde {\epsilon} - \epsilon ] \Big)
\sinh \Big( \frac{\pi \tau  }{2} [2eE_0f(z)\tau - \Tilde{\epsilon} + \epsilon ] \Big)}
    {
    \sinh\Big(\pi \tau \Tilde{\epsilon }\Big)
     \sinh\Big(\pi \tau \epsilon\Big) }
\end{equation}
\end{widetext}
and
\begin{equation}
    \Tilde{\epsilon}=\sqrt{m^2 + p_{\bot}^2 + \Big(p_z- 2\tau eE_0f(z)\Big)^2  }
\end{equation}
This result agrees with Ref \cite{Gies}. The arguments of the hyperbolic functions in \cref{Quantum_Vlasov} are large enough that we approximate the function $F$ as
\begin{equation}
    \label{Ftoexp}
    F\Big(\bp,E_0f(z),t\rightarrow \infty\Big) \approx 2 \,e^{\pi\tau \big(2\tau e E_0f(z)-\epsilon -\Tilde{\epsilon} \big)}
\end{equation}
The results \crefrange{Local_approx}{Ftoexp} will be used throughout the next subsection.

\begin{figure}
    \includegraphics[width=0.53\textwidth]{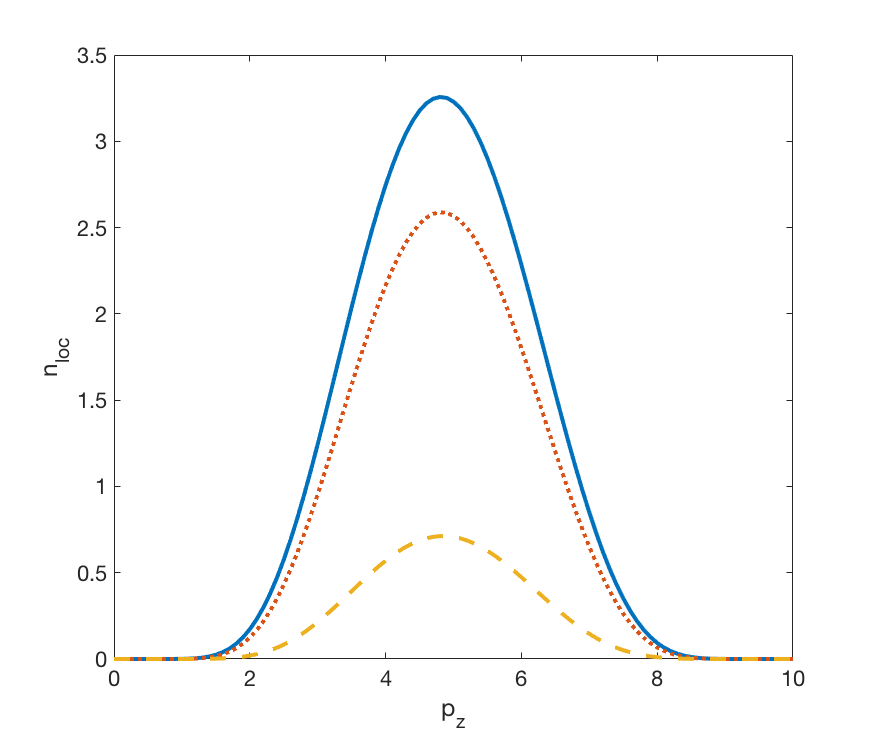}
    \caption{The number of pairs $\tilde{n}_\text{loc}(\bp,t\rightarrow \infty)$ as a function of the normalized parallel momentum $p_z/m$ for three different values of the normalized perpendicular momentum $p_{\bot}/m$, the solid curve has $p_{\bot}/m=0$, the dotted curve has $p_{\bot}/m=0.2$ and the dashed curve has $p_{\bot}/m=0.5$.}
\end{figure}

\subsection{The dependence on perpendicular momentum}

As seen from \cref{Quantum_Vlasov}, the perpendicular momentum only enters in
the equation system through the energy $\epsilon $. Consequently, the perpendicular
momentum has a limited effect on the basic physics of the problem, as
pointed out by e.g. Ref \cite{Gies} that wrote "It is known from the analysis of the Schwinger effect in spatially homogeneous electric fields that the orthogonal
momentum solely acts as an additional mass term and does
not change the qualitative behavior". Consequently Ref. \cite{Gies} put $p_{\perp}=0$ in their further analysis. This simplification can be further supported, by plotting the dependence of the pair production
rate on the \thinspace $p_{z}$ for different perpendicular momenta $%
p_{\bot }$. Considering the number of pairs  $\Tilde{n}_\text{loc}(\bp,t\rightarrow \infty)$ in \cref{Local_approx} where we use the configuration of the electric field in \cref{Electric_field}, the result is
displayed in Figure 1.
We can see that the production rate is diminished
with increasing $p_{\bot }$, just as if extra mass has been added to
the electrons and positrons. This indeed confirms the given motivations for
neglecting\thinspace\ the perpendicular momentum in the pair production
process. Particularly if the main aim is just to gain a qualitative
understanding for the dynamics. 

However, there are still a number of questions related to the perpendicular
momentum that need to be answered. For example, how does the full momentum
distribution $\Tilde{n}_\text{loc}(p_{z},p_{\bot })$ of the generated pairs look?
Importantly, depending on the magnitude of the perpendicular momentum, the
production rate can be more or less suppressed. Moreover, to what extent
does the over-estimation of the production  rate, introduced by omitting the
perpendicular momentum, depend on the parameters of the problem? In order to
answer these questions, we compute the full momentum distribution $%
\Tilde{n}_\text{loc}(p_{z},p_{\bot })$ from \cref{Local_approx}. 

In Figure 2, the distribution function  $\Tilde{n}_\text{loc}(p_{z},p_{\bot })$  is displayed for different magnitudes
of the electric field. \ As we can see, the contour curves are centered
around an average \thinspace value of $p_{z}$ that is higher for a stronger
electric field. Moreover, the characteristic spread in \thinspace $p_{\bot }$
and $p_{z}$ are both increasing with a stronger electric field. The
effective mass added in the production process is proportional to the
average value of  $p_{\bot }$, which in turn is proportional to the spread
in $p_{\bot }$ Since this is dependent on the magnitude of the electric
field, we can deduce that the error introduced by neglecting $p_{\bot }$ is
dependent on the magnitude of the electric field. In Figure 3 we have quantified this
observation by plotting $\Delta p$, the spread in $p_{\bot }$, as a function
of $E/E_{cr}$. Loosely equating $\Delta p$ with the added effective mass of
the pairs, gives a quick way to assess the accuracy in the common
approximation of dropping the dependence on $p_{\bot }$. In principle, the
spread in momentum also depend on the length of the pulse duration. However, the dependence on the pulse duration  is more or less negligible, and hence we omit plotting the result. 

A consequence of omitting the perpendicular momentum appears when studying the number density of produced pairs. For the general expression, we have
\begin{equation}
N=\int d^3p\, \Tilde{n}_\text{loc}(\bp,t\rightarrow \infty),
\end{equation}
and
we must use the simplified expression 
\begin{equation}
N_{\parallel}=\int dp_z\,\Tilde{n}_\text{loc}(p_z,t\rightarrow \infty)
\end{equation}
when there is no dependence on perpendicular momentum. However, the pair-production rate depends on the width of the distribution in perpendicular momentum space, which in turn depends on the magnitude of the electric field. As a result, the pair production rate $N_{\parallel}$ with the perpendicular momentum omitted, and the full expression  $N$ will scale differently with the electric field magnitude. In Fig. 4 we have studied this effect in the local density approximation using the same electric field profile as before. As can be seen, there is a general overestimation of the number of pairs using the approximation of parallel momentum only. To some extent the general overestimation could be fixed quite easily by introducing an overall correction factor in the evolution equation. However, for a self-consistent model with a dynamically varying electric field, we can not in general compensate for the fact that the overestimation of the produced pairs is dependent on the electric magnitude. As seen in Fig. 4, this overestimation is considerably larger for a weaker electric field. 

Naturally, more figures of the perpendicular momentum dependence can be produced. Still, the ones we have chosen should be enough to give a reasonable picture of the significance of the perpendicular momentum in
basic pair-production processes of the Schwinger-type.

\begin{figure}
\includegraphics[scale=0.18]{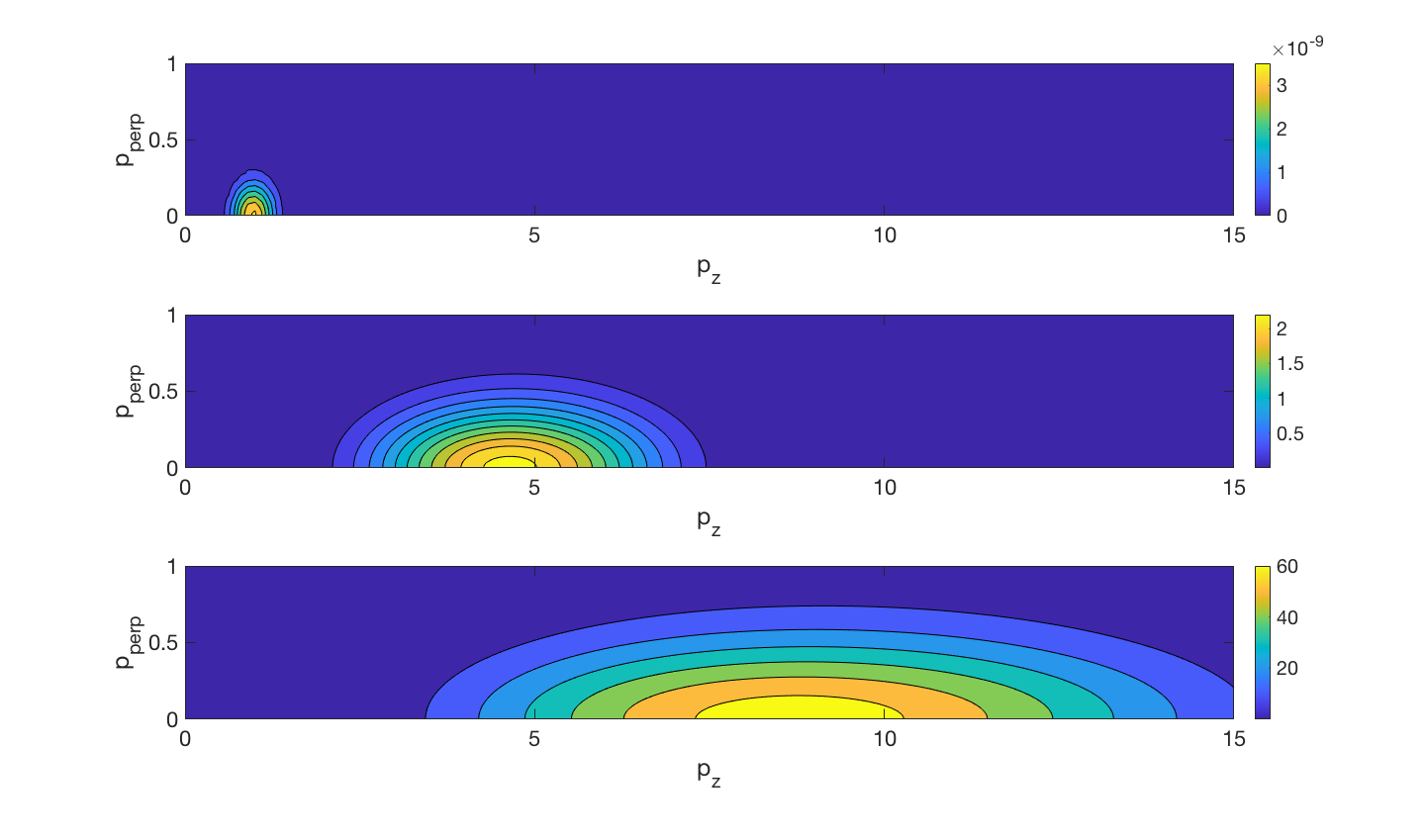}
\includegraphics[scale=0.18]{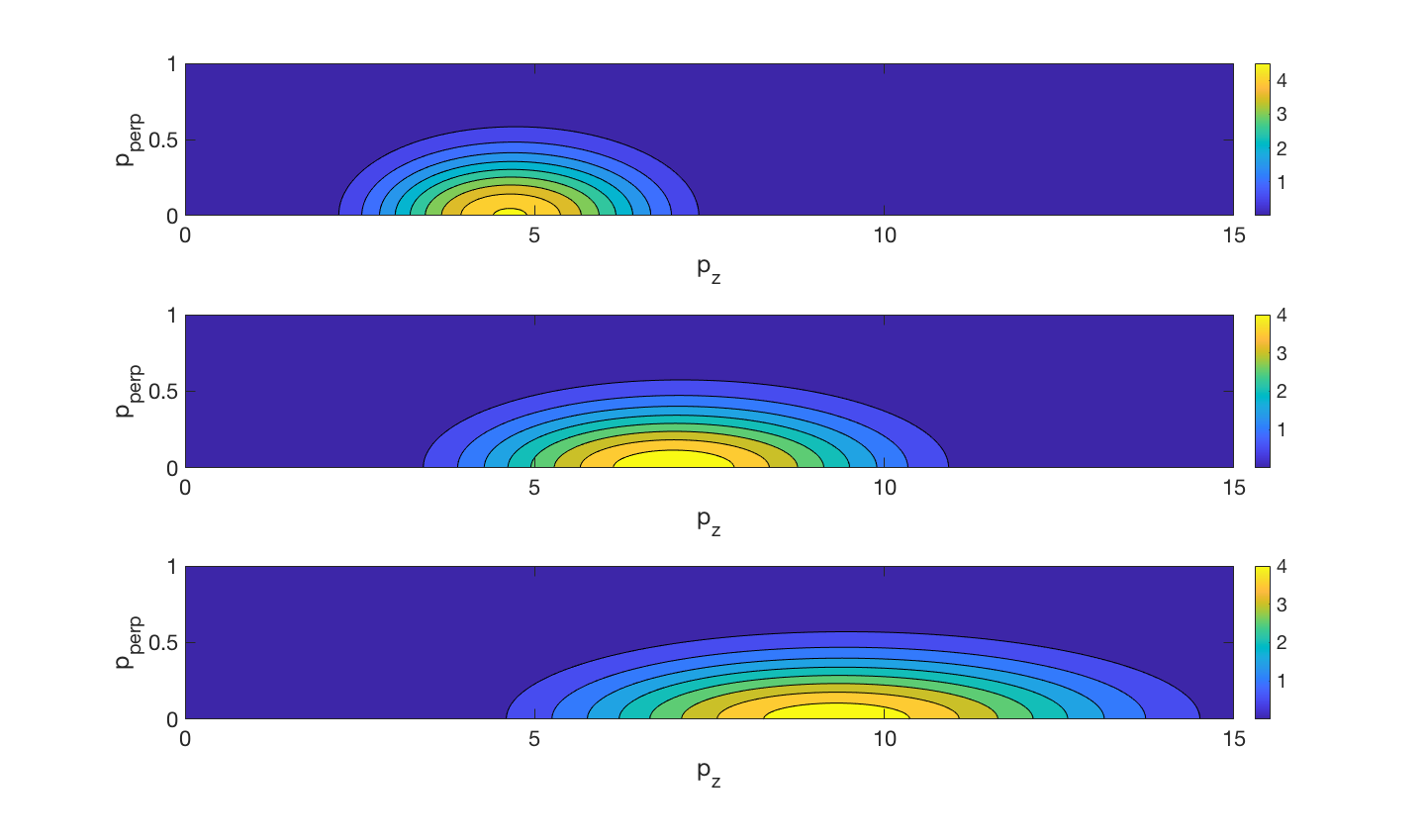}
\caption{Number of pairs $\Tilde{n}_\text{loc}(\bp)$ for: a) different amplitudes of the electric fields E=0.1,0.5,1 (the upper subfigure), b) different time duration $ \tau$=10,15,20 (the lower subfigure). }
\end{figure}
\begin{figure}
    \includegraphics[width=0.5\textwidth]{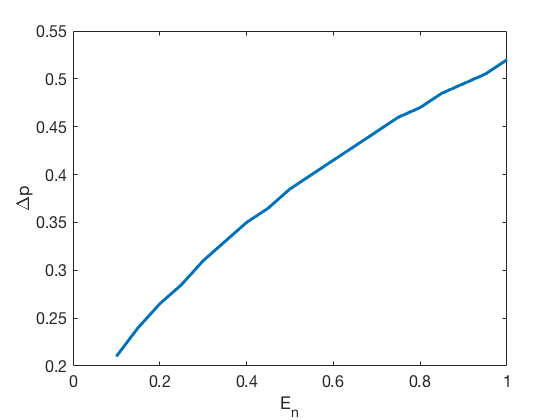}
    \caption{The spread of the the perpendicular momentum $\Delta p$ as a function of the normalized electric field $E_n=E/E_{cr}$.}
    \label{fig:my_label}
\end{figure}
\begin{figure}
    \includegraphics[width=0.5\textwidth]{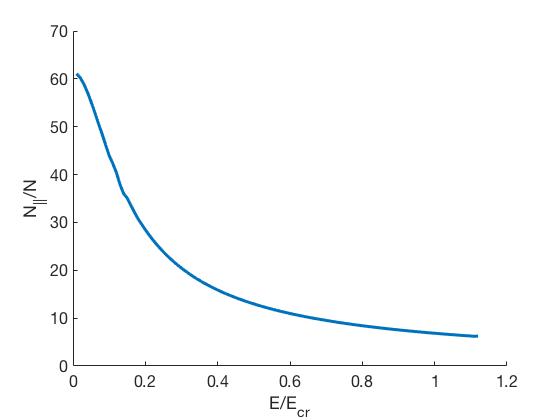}
    \caption{The fraction of the parallel number density $N_{\parallel}$ and the number density $N$ as a function of the normalized electric field $E/E_{cr}$. }
    \label{fig:my_label}
\end{figure}

\section{Summary and Discussion}
\label{Conclusion}
In this paper, we have studied the DHW-formalism in the 1D electrostatic limit. It turns out that for this case, the 16 scalar equations of the general theory can be reduced to four scalar equations given in (\ref{PDE_System}), which only needs to be complemented by Ampere's law (\ref{Ampers_law}). Systems similar to Eqs. (\ref{PDE_System}) have been studied previously, e.g. by Ref. \cite{Gies 2}, who, however, did not include the dependence on perpendicular momentum. While a perpendicular momentum dependence was included in Ref. \cite{Gies}, this paper  only studied the homogeneous limit. Also, none of these works treated the field self-consistently by simultaneously solving (\ref{Ampers_law}). 

To demonstrate the versatility of Eqs.~(\ref{PDE_System})-(\ref{Ampers_law}), we first applied the system to linearized electrostatic waves in plasmas.  The dispersion relation was derived, and shown to agree with well-known limiting cases. The issue of re-normalization, which is needed to treat the ultra-violet divergences associated with the vacuum background, is left for a future paper, however. In this context, it should be pointed out that a quantum-relativistic treatment of plasma waves is needed for very high plasma densities, such that the Fermi velocity is relativistic, as is the case for e.g. dense astrophysical objects.  

For problems of pair-production in a given field, it has been common to neglect the perpendicular momentum dependence, see e.g. Refs. \cite{Gies 2}. While this is a rather natural simplification, as the perpendicular momentum merely adds some extra mass to the pairs, nevertheless the accuracy of this approximation might not be very high. Studying Eqs.~(\ref{PDE_System}) for a given electric pulse with a temporal sech-profile, it is found that the approximation often is a useful one. Nevertheless, it is somewhat problematic to omit the perpendicular momentum dependence, as the error in the pair-production rate induced by this omission depends on the parameters of the problem. Specifically, for weakly inhomogeneous systems (such that the local density approximation is applicable), the perpendicular momentum of the generated pairs is close to linearly proportional to the electric field (cf. fig 3.) As a result, there is a general overestimation of the produced pairs when the perpendicular momentum is overlooked. While, in principle, a correction factor could be introduced to compensate for the overestimation, such a solution is not entirely satisfactory, as the correction factor would be dependent on the electric field magnitude, that could be varying dynamically in a self-consistent field model.  

The broader conclusion from the present study, is that the equation system (\ref{PDE_System})- (\ref{Ampers_law}) provides a useful basis for studying pair-creation in a plasma medium self-consistently. However, for field-strengths sufficiently high to give appreciable pair-production, the plasma dynamics will become strongly nonlinear. Thus, in order to study pair-production in a plasma, the analytic treatment of the present paper must be replaced by a numerical approach.


\begin{thebibliography}{99}
\bibitem{Temporal}
How slow the variations need to be for \cref{hartree} to apply is not obvious. This issue will be discussed in a forthcoming paper, where we will argue that for sufficiently intense fields, the temporal scale may even be allowed to approach and surpass the Compton scale.
\end{thebibliography}
\end{document}